\documentclass{modified_ws-procs9x6}

\newcommand{\zpr}{\mbox{$Z'$}}
\newcommand{\mzp}{\mbox{$M_{Z'}$}}
\newcommand{\upr}{\mbox{$U(1)'$}}

\newcommand{\be}{\begin{equation}}
\newcommand{\ee}{\end{equation}}
\newcommand{\bea}{\begin{eqnarray}}
\newcommand{\eea}{\end{eqnarray}}

\def\mxth{\mathsurround=0pt }
\def\xversim#1#2{\lower2.pt\vbox{\baselineskip0pt \lineskip-.5pt
  \ialign{$\mxth#1\hfil##\hfil$\crcr#2\crcr\sim\crcr}}}
\def\simgr{\mathrel{\mathpalette\xversim >}}
\def\simle{\mathrel{\mathpalette\xversim <}}

\begin{document}

\title{Big Bang Nucleosynthesis\\
Constraints on $Z'$ Properties\footnote{\uppercase{T}alk presented by \uppercase{H-S}. \uppercase{L}ee
at {\it \uppercase{SUSY} 2003:
\uppercase{S}upersymmetry in the \uppercase{D}esert}\/, 
held at the \uppercase{U}niversity of \uppercase{A}rizona,
\uppercase{T}ucson, \uppercase{AZ}, \uppercase{J}une 5-10, 2003.
\uppercase{T}o appear in the \uppercase{P}roceedings.}}

\author{Vernon Barger}

\address{Department of Physics, University of Wisconsin, Madison, WI 53706 USA}

\author{Paul Langacker}

\address{Department of Physics and Astronomy, University of Pennsylvania, Philadelphia, PA 19104 USA}

\author{Hye-Sung Lee}

\address{Department of Physics, University of Wisconsin, Madison, WI 53706 USA}


\maketitle

\abstracts{
In models involving new $TeV$-scale \zpr \ gauge bosons, the new \upr \ symmetry often prevents the generation of Majorana masses needed for a conventional neutrino seesaw, leading to three super-weakly interacting ``right-handed'' neutrinos, the Dirac partners of the ordinary neutrinos.
These can be produced prior to Big Bang Nucleosynthesis (BBN) by the
\zpr \ interactions, leading to a faster expansion rate and too
much $^4He$. We quantify the constraints on the \zpr \ properties
from nucleosynthesis for \zpr \ couplings motivated by a class of
$E_6$ models parametrized by an angle $\theta_{E6}$.
The decoupling temperature, which is higher than that of ordinary left-handed neutrinos due to the large \zpr \  mass, is calculated, and the equivalent number of extra weakly interacting neutrinos, $\Delta N_\nu$, is obtained numerically as a function of the \zpr \ mass, couplings, and the $Z$-$Z'$ mixing angle. The
$^4 He$ abundance from BBN gives the most stringent limit on $M_{Z'}$ unless $Z'$ coupling to the right-handed neutrinos are small.}

\section{An $E_6$-motivated \zpr \ model}
Many models from String theory or GUTs predict additional neutral gauge bosons $(Z')$.
$TeV$-scale $Z'$ models are particularly interesting since they can solve the $\mu$-problem in the Minimal Supersymmetric Standard Model (MSSM) in a natural way (by replacing $\mu {\hat H_1} \cdot {\hat H_2}$ with $h_s {\hat S} {\hat H_1} \cdot {\hat H_2}$).
The $E_6$ group involves two additional $U(1)$ factors when broken to the Standard Model (SM),$$E_6 \to SO(10) \times U(1)_\psi \to SU(5) \times U(1)_\chi \times U(1)_\psi.$$
A canonical $E_6$ GUT with a $TeV$-scale $Z'$  would lead to too rapid proton decay
mediated by $E_6$ exotics. However, the
 $E_6$ charge assignments for the ordinary and exotic particles provides a simple example
of an anomaly-free $U(1)'$ model which can be consistent with minimal gauge unification,
and serves as a prototype for (typically more complicated) charges from concrete string
constructions.

We assume that only one linear combination of the two extra $U(1)'$ symmetries survives at 
the $TeV$-scale.
The resultant charge will be a mixture of the two $U(1)'$ charges with a mixing angle
 $\theta_{E_6}$, $Q = Q_\chi \cos\theta_{E_6} + Q_\psi \sin\theta_{E_6}$.
We assume the gauge coupling constant $g_Z' \sim {\small \sqrt{\frac{5}{3}}} g_1$, which 
is motivated by minimal gauge unification with the MSSM gauge group.
Since the right-handed neutrino charge is in general non-zero, except for a particular
 value of $\theta_{E_6}$ at which the linear combination vanishes, any Majorana mass for the right-handed neutrinos is expected to be no larger than the $U(1)'$ breaking scale ($TeV$).
This is too small for the ordinary seesaw mechanism.
We simply assume that the neutrinos are Dirac particles with negligibly small mass 
by some mechanism, such as higher-dimensional operators or large extra dimension.

After electroweak and \upr \ symmetry breaking, two neutral massive gauge bosons, $Z$ and \zpr, can mix.
The  limit on this mixing angle is $|\delta| < (2-3) \times 10^{-3}$ from precision experiments,
while $M_{Z'} > (500 - 800)~ GeV$, depending on the model-dependent couplings and number of open
decay channels.

\section{Superweak right-handed neutrinos and BBN}
To find the  constraint on the \zpr \ mass
from BBN, we basically follow the approach of Steigman, Olive and Schramm \cite{steigman1,steigman2}.
In the early universe, $\Gamma(T)$ (the interaction rate of a particle $A$)
 is larger than $H(T)$ (the cosmological expansion rate), but as the universe expanded and cooled, $\Gamma$ became comparable to $H$ and particle $A$ decoupled at
 the decoupling temperature $T_d(A)$.

For the SM neutrinos, $\Gamma(T) \equiv n \left< \sigma v \right> \approx G_W^2 T^5$ with $G_W \propto { \frac{g_Z^2}{M_Z^2}}$.
Since the right-handed neutrinos are coupled only to \zpr,  which is much heavier than $Z$, we expect $G_{SW} \propto {\frac{g_Z'^2}{M_{Z'}^2}} \ll G_W$.
Since $\Gamma(T)$ of the right-handed neutrinos will be much smaller (super-weakly interacting) than other SM particles, we expect them to decouple much earlier (with higher  $T_d(\nu_R)$,
and their number densities at the time of BBN will be reduced relative to the ordinary
active neutrinos by the subsequent reheating of the latter by the annihilations of heavy
particles and by the effects of the quark-hadron phase transition. To determine $T_d$ it is
necessary to compute the interaction rates, expansion rates, and entropy at any given temperature,
taking into account the number and types (e.g., quarks or hadrons) of particles in equilibrium.

The interaction rate is computed using the cross-section 
for $\bar \nu_R \nu_R \to \bar f_i f_i$ which is, in the no-mixing and massless particles limit, 
$$\sigma \to N_C^i {\small \frac{s}{12 \pi}} \left( {\small \frac{g_Z'^2}{M_{Z'}^2}} \right)^2 Q(\nu_R)^2 \left( Q(f_{iL})^2 + Q(f_{iR})^2 \right),$$
confirming the previous expectation of $G_{SW}~ \propto~ \frac{g_Z'^2}{M_{Z'}^2}$.
For $T$ less than the quark-hadron transition temperature $T_c$, we replaced the final fermions with meson pairs.

On the other hand, during the radiation dominated epochs, $H(T) \propto \sqrt{G_N \rho_\gamma(T) g(T)}$
with photon energy density $\rho_\gamma(T) = a T^4$ and the effective number of degrees of freedom
$$g(T) = \sum_B g_B \left({\small \frac{T_B}{T}}\right)^4 + \sum_F {\small \frac{7}{8}} g_F \left({\small \frac{T_F}{T}}\right)^4.$$

The SM prediction for $g(T)$ at BBN $(T \approx 1~ MeV)$ is $43/4$ from  photons, electrons, positrons, and three SM neutrinos.
The additional right-handed neutrinos' contribution to the expansion rate can be parametrized by
the effective number  $\Delta N_\nu$ of additional neutrino types, where
$$ \Delta N_\nu  = \frac{3}{2} g_{\nu_R} \left({\small \frac{T_{\nu_R}}{T_{BBN}}}\right)^4,$$
where the 3 is the number of families, $g_{\nu_R}=2$ is the number of degrees of freedom,
and the $(T_{\nu_R}/T_{BBN})^4$ factor is from the dilution due to the reheating of ordinary 
neutrinos after $\nu_R$ decoupling.
In a typical analysis, $\Delta N_\nu \simle (0.3 - 1)$ from the observed $^4 He$ abundance
 $(\Delta Y \sim 0.013 \Delta N_\nu)$, which allows up to one 
 additional weakly interacting neutrino species.

Using entropy conservation in the early universe, one has
$$\Delta N_\nu = 3 \left({\small \frac{T_{\nu_R}}{T_{BBN}}}\right)^4 = 3 \left({\small \frac{g(T_{BBN})}{g(T_d (\nu_R) )}}\right)^{4/3},$$
where the $\nu_R$ are not included in the calculation of $g$.
We constrain $T_d (\nu_R)$ from the observed limit on $\Delta Y$. 
Then, using
$\Gamma(T_d(\nu_R)) = H(T_d(\nu_R))$, we obtain a lower limit on the \zpr \ mass.

\section{Numerical results and Conclusion}
\begin{figure}
\begin{center}
\includegraphics[height=40mm]{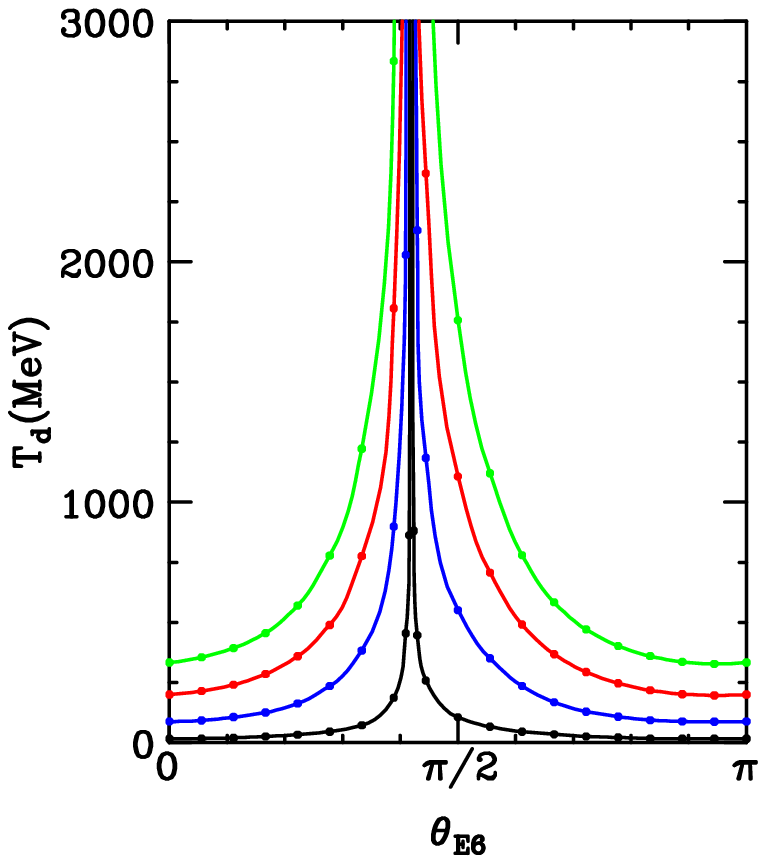}
\includegraphics[height=40mm]{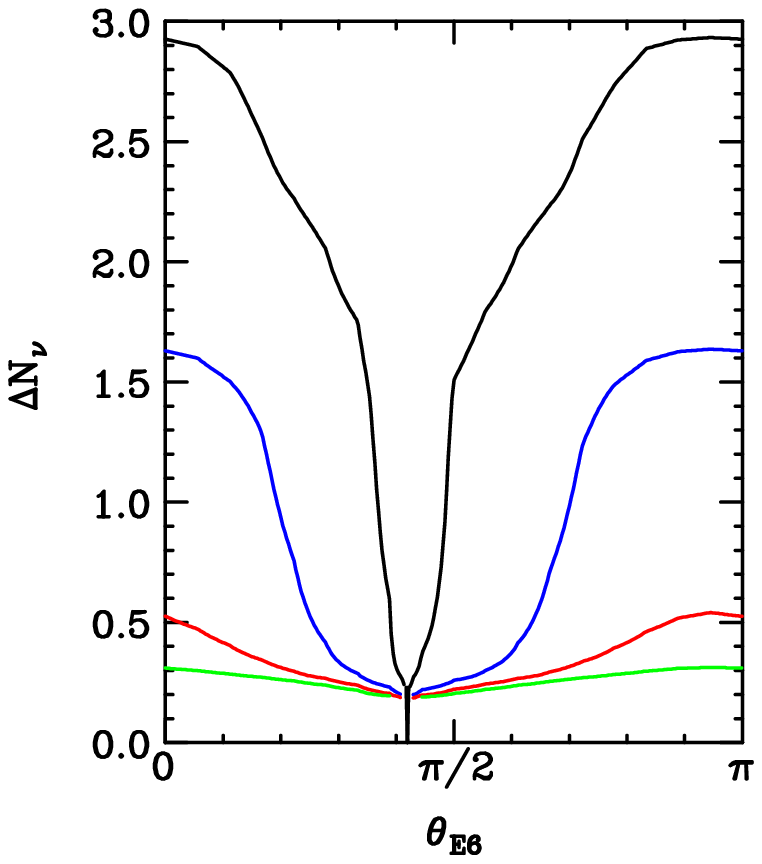}
\includegraphics[height=40mm]{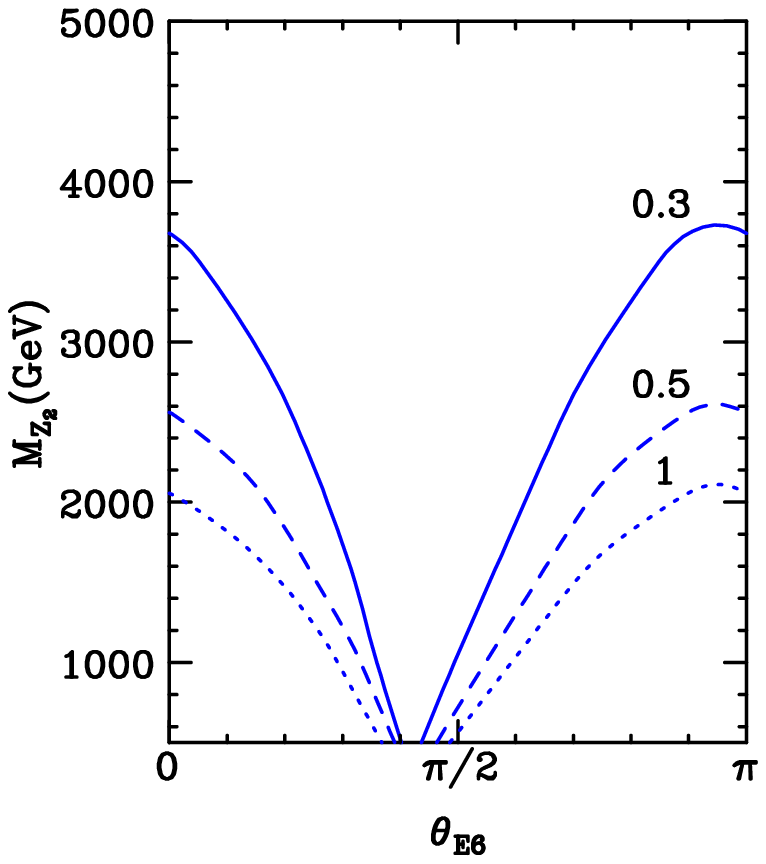}
\end{center}
\caption{$T_d$ and $\Delta N_\nu$ of the right-handed neutrinos (first two) and the $M_{Z'}$ limit for various limits on $\Delta N_\nu = 0.3,~ 0.5,~ 1$. No mixing and $T_c = 150~ MeV$ is assumed.}
\label{figure}
\end{figure}
In Figure \ref{figure} we show a special case in which there is no mixing between $Z$ and \zpr, and the quark-hadron transition temperature is $150~ MeV$.
(For more general cases, see~\cite{main}.)
In the first panel, as we increase the \zpr \ mass  from $0.5~ TeV$ to $1.5,~ 2.5,~ 3.5~ TeV$, the decoupling temperature of $\nu_R$ increases, as expected.
In the second panel, $\Delta N_\nu$ decreases because of the increasing decoupling temperature.
In the last panel  we plot the lower bound on $M_{Z'}$ for various upper limits on $\Delta N_\nu$.

At the singular point $\theta_{E_6} = 0.42 \pi$, the \upr \ charge for the $\nu_R$ vanishes and 
the $\nu_R$ decouples from  the \zpr , giving no contribution to $\Delta N_\nu$.
(A possible mechanism for obtaining this value is described in~\cite{sb}.)
Except around this point, BBN gives a much stronger bound on $M_{Z'}$ than present collider limits. (Mostly, \mzp $\simgr$ multi-$TeV$.)
For higher $T_c$, the constraint is even more severe. $(T_c$ is between $150$ and $400~ MeV.)$



\end{document}